\documentclass[prl,twocolumn,aps,showpacs]{revtex4}
\usepackage{graphicx}
\usepackage{bm}
\usepackage{amssymb} 
\begin{document}

\title{A simple microscopic description of quantum Hall transition without Landau levels}

\author{V. V. Mkhitaryan$^{1}$, V. Kagalovsky$^{2}$, and M. E. Raikh$^{1}$}

\affiliation{$^{1}$ Department of Physics, University of Utah, Salt Lake
City, UT 84112, USA\\
$^{2}$ Sami Shamoon College of Engineering, Beer Sheva 84100, Israel}
\begin{abstract}

By restricting the motion of high-mobility 2D electron gas
to a network of  channels with smooth confinement, we were able
to trace, both classically and quantum-mechanically, the interplay of
backscattering, and of the bending action of a weak magnetic field.
Backscattering limits the mobility, while bending initiates
quantization of the Hall conductivity.
We demonstrate that, in restricted geometry, electron motion
reduces to {\em two}  Chalker-Coddington networks,
with {\em opposite} directions of propagation along the
links, which are  weakly coupled by disorder.
Interplay of backscattering and bending
results in the quantum Hall transition in a {\em non-quantizing} magnetic
field, which decreases with increasing mobility. This is in accord with
scenario of floating up delocalized states.

\end{abstract}
\maketitle

\noindent{\em Introduction.}
Quantization of the Hall conductivity of a disordered 2D electron gas,
$\sigma_{xy}=n$,
 (in the units of $e^2/h$)
together with vanishing diagonal conductivity, $\sigma_{xx}$,
reflect the fact that in a perpendicular magnetic field
delocalized states always constitute a discrete set
\cite{Halperin82}.

\begin{figure}[b]
\centerline{\includegraphics[width=75mm,angle=0,clip]{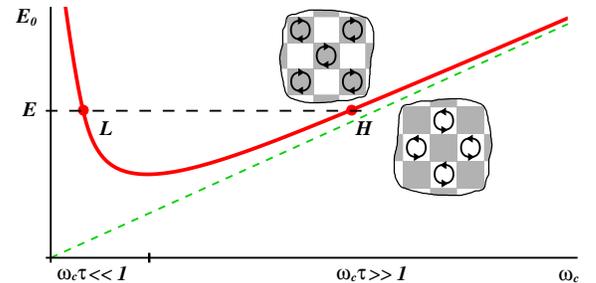}}
\caption{(Color online)  Energy position of the lowest delocalized
state, $E_0$, vs. magnetic field, $\omega_c$, as predicted in
Ref.~\onlinecite{Khmelnitskii}. For a given $E$, delocalization
transition occurs at points $L$ and $H$. Change of character of
electron motion near $H$ is illustrated schematically.
Corresponding change near $L$ is illustrated in Fig. \ref{diffph}
(a $\leftrightarrow$ e).} \label{levit}
\end{figure}

In a strong magnetic field, $\omega_c\tau \gg 1$, where $\omega_c$
is a cyclotron frequency and $\tau$ is the scattering time, energy
positions, $E_n$, of the delocalized states coincide with the
centers of well-resolved Landau bands. Such a strong-field limit
was the focus of theoretical studies of delocalization in a
magnetic field. Most appealing qualitative picture
\cite{Kazarinovetal} assumes a smooth disorder when the
eigenstates are well-defined Larmour circles drifting along
equipotential lines. Then delocalization corresponds to the {\em
classical} percolation threshold; localized states above $E_n$
are closed drift trajectories executed, {\em e.g.}, clockwise,
while the states below $E_n$ are closed drift trajectories
executed counter-clockwise, see Fig. \ref{levit}.

An alternative approach \cite{pru83}  to delocalization is based
on  renormalization-group equations, describing the evolution of
$\sigma_{xx}$, $\sigma_{xy}$ upon increasing the sample size, $L$,
\begin{eqnarray}
\label{pru1}
&&\frac{\partial\sigma_{xx}}{\partial\ln L}=-\frac{1}{2\pi^2\sigma_{xx}}-
\sigma_{xx}^2De^{-2\pi\sigma_{xx}}\cos(2\pi\sigma_{xy}),\\
&&\frac{\partial\sigma_{xy}}{\partial\ln L}=
-\sigma_{xx}^2De^{-2\pi\sigma_{xx}}\sin(2\pi\sigma_{xy}),\label{pru2}
\end{eqnarray}
where $D$ is a dimensionless constant. First term of Eq.
(\ref{pru1}) originates from {\em interference} of electron
multiple-scattering paths: two paths corresponding to the {\em
same} scatterers but different sequences of scattering events
interfere even in the presence of Aharonov-Bohm phases. Second
term reflects the {\em orbital} action of magnetic field: by
curving electron trajectories it tends to destroy the
interference. When the ``phase'' and ``orbital'' terms compensate
each other, delocalization transition takes place.

Field-theoretical approach \cite{pru83} yields a highly nontrivial
prediction first pointed out by Khmelnitskii \cite{Khmelnitskii}.
Namely, solving Eqs. (\ref{pru1}), (\ref{pru2}) together with
classical initial condition
$\sigma_{xy}(\omega_c)=\sigma_0\omega_c\tau(1+\omega_c^2\tau^2)^{-1}$,
where $\sigma_0 \propto E_n$ is the dimensionless conductance at
$\omega_c=0$, yields $E_n =\hbar\omega_c\left(n+\frac{1}{2}\right)
\left[1+(\omega_c\tau)^{-2}\right]$. As shown in Fig.~\ref{levit}
 for $n=0$, the high-field part, $\omega_c\tau\gg 1$,
of $E_0$ follows the center of the lowest Landau level,
while the low-field part ``floats up'' as $\omega_c\tau \rightarrow 0$.
This prediction is essential component of the global phase
diagram \cite{Kivelson}.

Qualitative classical picture \cite{Kazarinovetal} applies to the
high-field part and illustrates the restructuring, see Fig.
\ref{levit}, of the motion of guiding center, which accompanies
the crossing of $E_0$  by the Fermi level,
upon increasing
magnetic field ($1 \rightarrow 0$ transition into the quantum Hall insulator
\cite{Pryadko}).

While there is certain experimental evidence \cite{Exp1, Exp2,
Exp3, Exp4, Exp5, Exp6, Exp7}  that floating up of $E_0(\omega_c)$
indeed takes place, tight-binding numerical studies \cite{Yang,
Wen, Sch} are less conclusive. There is a fundamental reason
\cite{Huckestein} for this lack of conclusiveness. Indeed,
significant floating up occurs for large
$\sigma_0>(\omega_c\tau)^{-1}\gg 1$. For such $\sigma_0$, upon
moving along the dashed horizontal line in  Fig. \ref{levit} the
localization length undergoes a steep growth from orthogonal value
$\xi_o \sim \exp(\pi\sigma_0)$ to the unitary value $\xi_u
\sim\exp(\pi^2\sigma_0^2)$ already in very small fields
$\omega_c\tau\simeq \sigma_0^{-1} e^{-2\pi\sigma_0}$, way before
the expected transition at point $L$. Such large $\xi_u$ is a
major obstacle for numerics to capture the descending region of
the line of transitions in Fig. \ref{levit}. Accessing this region
requires to construct an efficient minimal description of a
weak-field transition, as transparent as the picture
\cite{Kazarinovetal} sketched near the point $H$ in
Fig.~\ref{levit}. This goal is achieved in the present paper. The
key step of our construction is separation of the spatial regions
with disorder-induced scattering and  field-induced bending.

\begin{figure}[t]
\centerline{\includegraphics[width=80mm,angle=0,clip]{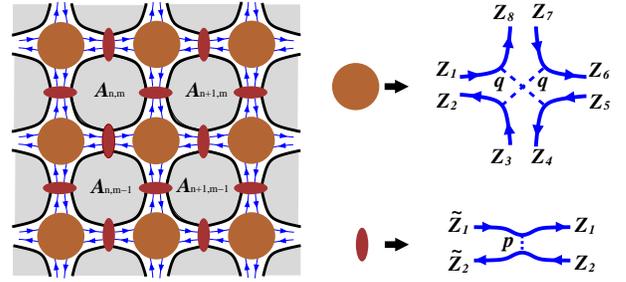}}
\caption{(Color online) Left: Restricted electron motion over
point contacts and bend-junctions is illustrated; $A_{n,m}$ are
the centers of forbidden regions. Right: Scattering matrices of
the junction and of the point contact. } \label{network}
\end{figure}

\noindent{\em Restricted electron motion.} ({\em i}) In contrast
to unidirectional motion in strong fields, we allow
counter-propagating paths in the regions where {\em orbital}
action of magnetic field is negligible. We achieve this by
restricting electrons to narrow point contacts, see Fig.
\ref{network}. At the same time, we assume that the {\em phase}
action of magnetic field is well-developed in each point contact, i.e.,
the area of the contact is threaded by many flux quanta. Presence
of disorder is incorporated by allowing mutual backscattering of
two counterpropagating waves. We quantify the strength of
backscattering with probability, $p$, so that the scattering
matrix of the contact has the form
\begin{equation}
\label{Matrixp}
\left(
\begin{array}{c}
Z_1\\
\tilde{Z}_2
\end{array}\right)=\left(\begin{array}{cc}
{\scriptstyle \sqrt{1-p}}&{\scriptstyle \sqrt{p}}\\
{\scriptstyle -\sqrt{p}}&{\scriptstyle \sqrt{1-p}}
\end{array}\right) \left(
\begin{array}{c}
\tilde{Z}_1\\
Z_2
\end{array}\right),
\end{equation}
with amplitudes $Z_i$, $\tilde{Z}_i$, Fig. \ref{network}, having
random phases.

\begin{figure}[b]\vspace{-0.3cm}
\centerline{\includegraphics[width=90mm,angle=0,clip]{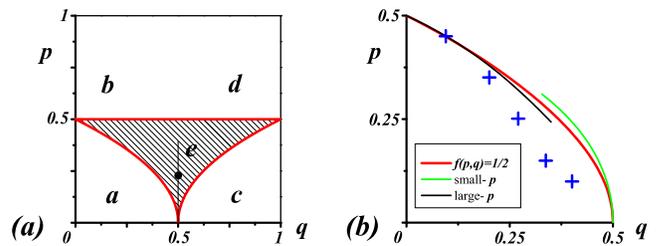}}
\caption{(Color online) ({\em a}) Classical metallic phase, dashed
region (e) in the $(p,q)$-plane separates different insulating
phases of the model Fig. \ref{network}; ({\em b}) Line of
classical delocalization transitions in the $(p,q)$-plane.
Asymptotes Eqs. (\ref{smallp}), (\ref{largep}), are shown with
green and black lines, respectively; red curve interpolates
between the two regimes. Crosses show the positions of quantum
delocalization transition inferred from simulations for five
values of ``energies'', $p$. Quantized $\sigma_{xy}$ is $1$ in
phase (a), $-1$ in phase (c), and $0$ in (b), (d), and (e).}
\label{phasediag}
\end{figure}

\noindent({\em ii}) The orbital action of magnetic field takes
place in the junctions between the point contacts,
Fig.~\ref{network}. To simplify the description of the junction,
we assume that an electron incident, say, from the left, after
several bounces \cite{Scat}  off the walls exits  either ``up'' or
``down'', i.e., both forward and backward scattering channels are
suppressed. This assumption allows us to quantify the bending
strength of the junction by a single parameter, $q$, the
deflection probability to the right, Fig. \ref{network}. Then the
deflection probability to the left is $(1-q)$.  The Hall
resistivity of the junction \cite{Scat} is then given by
$R_H=(e^2/h)[q-(1-q)]/[q^2+(1-q)^2]$, so that $q=1/2$ corresponds
to a zero field. Expression for the scattering matrix of the
junction is the following
\begin{equation}
\label{Matrixq}
\left(
\begin{array}{c}
Z_2\\
Z_4\\
Z_6\\
Z_8
\end{array}\right)=\left(
\begin{array}{cccc}
0&{\scriptstyle\sqrt{1-q}}&0&{\scriptstyle \sqrt{q}}\\
{\scriptstyle \sqrt{q}}&0&{\scriptstyle \sqrt{1-q}}&0\\
0&{\scriptstyle -\sqrt{q}}&0&{\scriptstyle \sqrt{1-q}}\\
{\scriptstyle \sqrt{1-q}}&0&{\scriptstyle -\sqrt{q}}&0
\end{array}\right)\left(
\begin{array}{c}
Z_1\\
Z_3\\
Z_5\\
Z_7
\end{array}\right).
\end{equation}
With scattering matrices Eqs. (\ref{Matrixp}), (\ref{Matrixq})
defined, the problem of electron localization by disorder in a
magnetic field reduces to the effective network model, which can
be studied by transfer-matrix method, similar to
Chalker-Coddington (CC) model \cite{CC}, which describes
delocalization transition at the point $H$ in Fig.~\ref{levit}. As
in Ref. \onlinecite{CC}, ``unitary'' disorder  is incorporated via
random phases of the link amplitudes, $Z_i$. Delocalization
transitions in the network define a line on the $(p,q)$ plane.
Important is that this line can be converted into the dependence
$E_0(\omega_c)$. Indeed, parameter
 $q$ reflects the strength of magnetic field, so that
$\left(\frac{1}{2}-q \right) \propto \omega_c$, while the
backscattering probability, $p$, decreases monotonously
with increasing energy. Thus, the floating scenario
is equivalent to the statement that $p(q)$-line approaches
$p=0$ as $q$ approaches $\frac{1}{2}$.
Below we argue that the form of  $p(q)$-line of
delocalization transitions
is the one shown in Fig. \ref{phasediag}(a) (region $q<\frac12$),
so that it indeed yields the dependence, $p(q)$,  corresponding
to the floating of $E_0(\omega_c)$.

In the CC model the transmission of the nodes with ``height'',
$E_i$, and ``width'', $\Gamma$, is given by the Fermi function
\cite{Fertig}
$T(E,E_i)=\left\{1+\exp\left[(E_i-E)/\Gamma\right]\right\}^{-1}$.
Qualitative strong-field picture of the transition
\cite{Kazarinovetal} emerges when the spread, $W$, of heights,
$E_i$, is $\gg \Gamma$. Then the quantum interference can be
neglected up to large distances, ${\cal R}_{cl}=
\left(W/\Gamma\right)^{4/3}\gg 1$, determined by the classical
percolation. For smaller distances, one can replace $T(E,E_i)$ by
a step-function, $\Theta(E_i-E)$. Adopting the same approach, we
assume that ({\em i}) in Fig. \ref{network} full transmission
takes place in  $p$ percent of point  contacts, and full
reflection  in the rest $(1-p)$ percent; ({\em ii}) a given
junction deflects only to the left in $q^2$ percent of cases, only
to the right in $(1-q)^2$ percent of cases; in the remaining
$2q(1-q)$ percent the deflection takes place both to the left and
to the right depending on incoming channel.

\begin{figure}[b]
\centerline{\includegraphics[width=80mm,angle=0,clip]{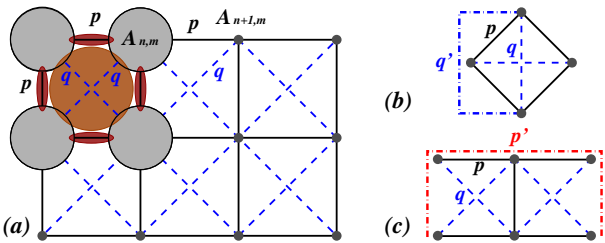}}
\caption{(Color online) ({\em a}) In the classical limit, electron
motion reduces to the  bond percolation on the lattice consisting
of p and q -bonds; ({\em b}) and ({\em c}) are graphic
representation of Eqs. (\ref{smallp}) and (\ref{largep}),
respectively.}
 \label{bondperc}
\end{figure}

\noindent{\em Phase diagram}. The key observation that allows to
establish the phase diagram Fig. \ref{phasediag} is that the
classical electron motion over the lattice of point contacts and
junctions can be reduced to a  single problem of {\em joint} bond
percolation over ``p'' and ``q'' -bonds. To substantiate this
statement, we focus on the grey squares in Fig.~\ref{network},
which are ``forbidden'' regions for electrons, and notice that
electron scattering processes {\em both} in point contacts and
junctions effectively establish bonds between these regions. More
specifically, if electron is backscattered in a point contact, we
consider that the {\em centers} $A_{n,m}$ and $A_{n,m-1}$ of the
squares, adjacent to this contact, are connected by a bond, see
Fig.~\ref{bondperc} . Further, if electron is bent-scattered by a
junction, say, in the direction left $\rightarrow$ down, we
identify this process with establishing a bond between the centers
of the squares $A_{n,m}$ and $A_{n-1,m-1}$. The above
identification reduces the classical motion through the network
with sites in the centers of squares,  see  Fig.~\ref{bondperc}.

\begin{figure}[b]
\centerline{\includegraphics[width=75mm,angle=0,clip]{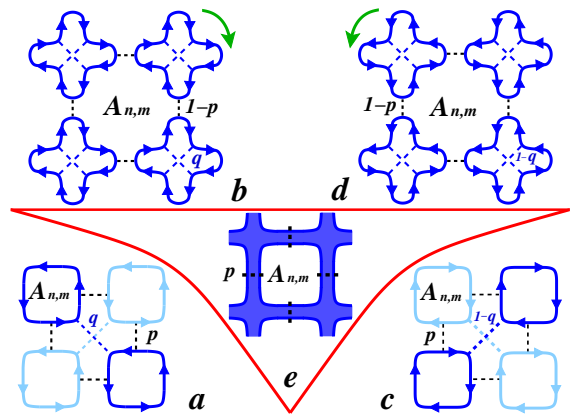}}
\caption{(Color online) Illustration of the structure of five
different phases in Fig. \ref{phasediag}. Electron moves
predominantly along solid directed lines. In phases {\em a} and
{\em c} the motion is either within dark or within light
subnetwork. In phase {\em e} coupling of subnetworks by p -bonds
opens a metallic phase.}
 \label{diffph}
\end{figure}

\noindent{\em Structure of phases.}  We start from the region of
strong reflection, $p>\frac12$. Counterpropagating
channels in point contacts, Fig. \ref{network}, are essentially
``short-circuited''. Hence,  no delocalization occurs upon
increasing magnetic field, $\left(\frac12 -q\right)$. Localized
states are illustrated schematically in Fig. \ref{diffph} in the
limit $(1-p)\ll 1$. It is apparent that crossover from clockwise
rotation (b) to counterclockwise rotation (d) upon passing the
zero-field line $q=\frac12$ takes place without delocalization.
This absence of delocalization is consistent with low-energy part
of Fig. \ref{levit}, because  $p> \frac12$ corresponds to small
$\sigma_0$. Along the line $p=\frac12$ the p -bonds alone
constitute a critical network. It is seen from Fig. \ref{diffph}
that, as $p$ is  reduced below $\frac12$, q -bonds sustain
percolation, manifesting that metallic behavior for $\sigma_0>1$
persists up to strong magnetic fields, $\omega_c\tau \sim 1$.

We now turn to the most interesting region of small $p$ (high
energies in Fig. \ref{levit}). In this domain the overall
connectivity of the network is dominated by the q -bonds.
Moreover, at $p=0$ the light-blue  and  dark-blue subnetworks are
{\em completely decoupled}. Small finite $p$ becomes essential in
the vicinity of $q=\frac12$, when both subnetworks of q -bonds are
critical. Now even weak non-zero $p$, by coupling the subnetworks,
results in opening of a {\em metallic} region (e) in the phase
diagram. This disorder-induced coupling of critical subnetworks is
the fundamental underlying mechanism for the restructuring of
states near the point $L$, Fig. \ref{levit}. In effect,
transformation  (e) $\rightarrow$ (a) with increasing field,
$\left(\frac12 -q\right)$, is the counterpart of transformation
near the point $H$ in quantizing $\omega_c$.

Fig.  \ref{diffph} (a) also illustrates that at small $p$ both
subnetworks are chiral. Transformation into the phase (e) upon
decreasing magnetic field is accompanied by the change of the Hall
conductivity from quantized to finite value smaller than $1$. Full
suppression of $\sigma_{xy}$ in the region (e) occurs only when
interference drives this region into the Anderson insulator, so
that the difference between (e) and strongly localized phases (b)
and (d) vanishes.

Calculation results for the phase boundaries are shown in Fig.
\ref{phasediag}(b) only for $q<\frac{1}{2}$ due to $q \rightarrow
(1-q)$ duality \cite{footnote}. The end-points
$(p,q)=\left(0,\frac{1}{2}\right)$ and
$(p,q)=\left(\frac{1}{2},0\right)$  are conventional bond
percolation thresholds for percolation  over regions $A_{m,n}$,
connected by q -bonds and by p -bonds, respectively, see Fig.
\ref{bondperc}. Next, consider small $p\ll 1$ and suppose that $q$
is slightly smaller than $\frac{1}{2}$. Then, instead of missing q
-bond, a {\em pair of one horizontal and one vertical} p -bonds
can provide the connection (there are two such variants, see Fig.
\ref{bondperc}b). Resulting shift of the threshold position is
determined by the condition
\begin{equation}\label{small}
f(p,q)=q^{\prime}=q+2(1-q)p^2(1+q) =1/2,
\end{equation}
where $q^{\prime}$ is the probability that effective q -bond
connects.

When $q$ is small, the role of q -bonds is to promote  percolation
over p -bonds. Unlike the previous case, the shift,
$(\frac{1}{2}-p)$, of the threshold is {\em linear} in $q$. This
is because in certain rare cases {\em one} q -bond takes on the
role of a missing p -bond. Assume that a vertical p -bond is
absent while one of the neighboring horizontal p -bonds is
present. Then the connection between the ends of the missing bond
can be established in two steps: first via this horizontal p -bond
and then via a q -bond, Fig.  \ref{bondperc}. Quantitatively, the
boundary $p(q)$ at small $q$ can be obtained from the real-space
renormalization group procedure \cite{Reynolds}. In Ref.
\onlinecite{Reynolds} the probability that the superbond,
illustrated in Fig.~\ref{bondperc}c connects, is given by
$f_0(p)=p^5+5p^4(1-p)+8p^3(1-p)^2+2p^2(1-p)^3$, where the last
three terms correspond to realizations when superbond connects
with one, two, and three original bonds removed; $f_0(p)=\frac12$
yields the exact threshold $p=\frac{1}{2}$. Using the fact that q
-bonds can restore the connectivity, and selecting suitable
realizations out of all $2^4$ possible states of q -bonds amounts
to the following modification of the probability,
$p^{\prime}=f(p,q)$, that superbond  connects
\begin{eqnarray}
\label{largep}
&&f(p,q)=f_0(p)+10p^2(1-p)^3\bigl[2q-q^2\bigr]\\
&&+p(1-p)^4\bigl[4q+8q^2-8q^3+q^4\bigr]+
(1-p)^5\bigl[2q^2-q^4\bigr].\nonumber
\end{eqnarray}
Upon equating $f(p,q)$ to $\frac{1}{2}$, Eq. (\ref{largep}) yields
the boundary of percolation transition at small $q$.

\noindent{\em Localization length.} The boundary {\em a}-{\em e}
of ``classical'' phase diagram Fig. \ref{phasediag}, is
characterized by  the critical exponent $\nu=\frac43$.
Simulations, see below, indicate that, with quantum interference,
the phase diagram Fig. \ref{phasediag}b remains unchanged, while
metallic phase in Fig. \ref{diffph}e turns into the Anderson
insulator with ``unitary'' localization length, $\ln \xi_u
=\pi^2\sigma_0^2$, where the Drude conductance is related to $p$
as $p=c\sigma_0^{-1}$, with $c \sim 1$. Interference also modifies
the divergence of the localization length
\begin{equation}
\label{xi} \xi(p,q) \sim |f(p,q)- 1/2|^{-\nu}
=|3p^2/2-\left(1/2-q\right)|^{-\nu},
\end{equation}
in the neighboring insulating region,  Fig. \ref{diffph}a, by
changing $\nu$ from $\nu=\frac{4}{3}$, to quantum $\nu \approx
\frac{7}{3}$. Starting from small field, $\left(\frac12
-q\right)\ll p^2$, $\xi$ retains its value $\xi_u \propto
\exp(\pi^2c^2/p^2)$ up to the narrow vicinity of the boundary,
$\left(\frac12 -q\right)=\frac32p^2$, when it crosses over to
diverging Eq. (\ref{xi}). The width of the ``quantum'' region can
be estimated as
\begin{equation}
\label{width}
\frac{\delta\omega_c}{\omega_c}=
\frac{\delta\left(
1/2-q\right)}{\left(1/2-q\right)}
=\frac1{p^2}e^{-\frac{3c\pi^2}{7p^2}}
=\frac{\sigma_0^2}{c^2}e^{-\frac{3}{7}\pi^2\sigma_0^2}.
\end{equation}
This region rapidly narrows in course of floating up.

\noindent{\em Quantum treatment of the network.} Numerical
simulations, employing matrices Eqs. (\ref{Matrixp}),
(\ref{Matrixq}) for nodes and incorporating random phases into the
link amplitudes, are required to verify the above predictions
based on the classical picture. They are also supposed to verify
that domain (e) in  Fig. \ref{phasediag} is, quantum-mechanically,
insulating. The results of transfer-matrix analysis  of the
two-channel \cite{Kagalovsky} network Fig. \ref{network} are shown
in  Fig. \ref{phasediag}b for five values of ``energy'', $p$.
Usual simulation procedure \cite{KramerMac}  was employed: upon
constructing a transfer matrix of a slice ($M$ nodes in transverse
direction with periodic boundary conditions) the net transfer
matrix of a system of length, $N$ (typical $N=240000$), was
obtained and diagonalized, yielding the Lyapunov exponents,
$\lambda_i(p,q)$, related to the eigenvalues as $\exp (\lambda_i
N)$. Localization length, $\xi_M(p,q)$, was inferred from the
smallest positive exponent: $\xi_M(p,q)=\lambda_{M/2}^{-1}$.
Simulation  confidently confirm that for finite ``energies'',
$p>0$, quantum system is insulating at $q=0.5$, while the state
with $p=0$ and $q=0.5$ is extended. As seen in
Fig.~\ref{phasediag}b, the discrepancy between classical and
quantum treatments is small.

Note in conclusion that among various network models studied
\cite{Kramer}, the closest to ours is the model \cite{Bocquet}.
Unlike Ref.~\onlinecite{Bocquet} our Eq.~(\ref{Matrixq}) describes
scattering, say, to the right, with the {\em same} probability,

\noindent{\em Acknowledgments.} This work was supported by the BSF
grant No. 2006201.

\end{document}